\documentclass[conference]{IEEEtran}

\usepackage{amsmath}
\usepackage{amsfonts}
\usepackage{amssymb}
\usepackage{epsf}
\usepackage{cite}
\usepackage{balance}
\usepackage{fancyhdr}
\usepackage{graphicx}
\usepackage{subfigure}
\usepackage[blocks]{authblk}
\usepackage[stable]{footmisc}
\usepackage{float}
\usepackage{fancyhdr}

\makeatletter
\def\ps@IEEEtitlepagestyle{%
  \def\@oddfoot{\mycopyrightnotice}%
  \def\@evenfoot{}%
}
\def\mycopyrightnotice{%
  {\footnotesize \copyright 2015 IEEE. Personal use of this material is permitted. Permission from IEEE must be obtained for all other uses, in any current or future media\hfill}
  \gdef\mycopyrightnotice{}
}

\newcounter{subeq}
\renewcommand{\thesubeq}{\theequation\alph{subeq}}
\newcommand{\newsubeqblock}{\setcounter{subeq}{0}\refstepcounter{equation}}
\newcommand{\mysubeq}{\refstepcounter{subeq}\tag{\thesubeq}}
\IEEEoverridecommandlockouts
\date{}
\begin{document}
\title{Error Performance Analysis of FSO Links with Equal Gain Diversity Receivers over Double Generalized Gamma Fading Channels}

\author{Mohammadreza Aminikashani}
\author{Mohsen Kavehrad}

\affil{Department of Electrical Engineering\authorcr
The Pennsylvania State University, University Park, PA 16802\authorcr
Email: \{mza159, mkavehrad\}@psu.edu\authorcr}

\maketitle
\begin{abstract}
Free space optical (FSO) communication has been receiving increasing attention in recent years with its ability to achieve ultra-high data rates over unlicensed optical spectrum. A major performance limiting factor in FSO systems is atmospheric turbulence which severely degrades the system performance. To address this issue, multiple transmit and/or receive apertures can be employed, and the performance can be improved via diversity gain. In this paper, we investigate the bit error rate (BER) performance of FSO systems with transmit diversity or receive diversity with equal gain combining (EGC) over atmospheric turbulence channels described by the Double Generalized Gamma (Double GG) distribution. The Double GG distribution, recently proposed, generalizes many existing turbulence models in a closed-form expression and covers all turbulence conditions. Since the distribution function of a sum of Double GG random variables (RVs) appears in BER expression, we first derive a closed-form upper bound for the distribution of the sum of Double GG distributed RVs. A novel union upper bound for the average BER as well as corresponding asymptotic expression is then derived and evaluated in terms of Meijer’s G-functions.
 \end{abstract}

\section{Introduction}\label{INTRODUCTION}
\IEEEPARstart{F}{ree-space} optical (FSO) communication, which has been receiving growing attention, refers to terrestrial line-of-sight optical transmission through the atmosphere using lasers or light emitting diodes (LEDs). FSO systems offer cost-effective, license-free and high capacity communication which is a promising solution for the "last mile" problem. The unique properties of these systems also make them appealing for a number of other applications including fiber backup, wireless metropolitan area network extensions, broadband access to remote or underserved areas, and disaster recovery.\cite{1,2,kashani2}.

A major performance impairing factor in FSO systems is the effect of atmospheric turbulence-induced fading resulting in random fluctuations in the received signal \cite{3}. Atmospheric turbulence occurs as a result of the variations in the refractive index due to inhomogenities in the temperature and the pressure of the atmosphere. In the literature, many statistical models have been proposed to model this random phenomenon for different degrees of turbulence severity. Under weak turbulence conditions, log-normal distribution has been widely used to model the random irradiance fluctuations \cite{4,5,6,6.1,6.2,kashani3}. As the strength of turbulence increases, log-normal distribution shows large deviations from the experimental data. Therefore, other statistical models such as K \cite{7}, I-K \cite{8}, log-normal Rician \cite{9}, Gamma-Gamma \cite{10}, M \cite{11} and Double Weibull \cite{12} distributions have been proposed to cover a wide range of turbulence conditions. Recently, a unifying statistical distribution named Double Generalized Gamma (Double GG) was proposed in \cite{kashani} which is valid under all range of turbulence conditions and contains most of the existing models in the literature as special cases.

Spatial diversity techniques provide a promising approach to mitigate turbulence-induced fading. Over the years, the performance of FSO systems with spatial diversity have been extensively studied over the most commonly utilized turbulence-induced fading models including log-normal, K, negative exponential and Gamma-Gamma \cite{15,20,21,Farid, Antonio}. In \cite{ICC}, a unified closed-form expression for the BER of single-input multiple-output (SIMO) FSO systems with optimal combining (OC) receiver over Double GG channel was proposed. However, the BER performance of equal gain combining (EGC) receivers, which is equivalent to that of multiple-input single-output (MISO) FSO systems, was presented in integral form. In fact, the main difficulty in studying EGC receivers is the distribution of the sum of Double GG R.Vs required to be derived in order to obtain a closed-form solution.

\begin{figure*}[t]
\begin{equation}\label{eq1}
{{f}_{{{I}_{n}}}}\left( {{I}_{n}} \right)=\frac{{{\gamma }_{2,n}}{{p}_{n}}{{p}_{n}}^{{{m}_{2,n}}-1/2}{{q}^{{{m}_{1,n}}-1/2}}{{I}^{-1}}}{{{\left( 2\pi  \right)}^{{\left( {{p}_{n}}+{{q}_{n}} \right)}/{2}\;-1}}\Gamma \left( {{m}_{1,n}} \right)\Gamma \left( {{m}_{2,n}} \right)}G_{{{p}_{n}}+{{q}_{n}},0}^{0,{{p}_{n}}+{{q}_{n}}}\left[ {{\left( \frac{{{\Omega }_{2,n}}}{{{I}^{{{\gamma }_{2,n}}}}} \right)}^{{{p}_{n}}}}\frac{p_{n}^{{{p}_{n}}}q_{n}^{{{q}_{n}}}\Omega _{1,n}^{{{q}_{n}}}}{m_{1,n}^{{{q}_{n}}}m_{2,n}^{{{p}_{n}}}}|\begin{matrix}
   \Delta \left( {{q}_{n}}:1-{{m}_{1,n}} \right),\Delta \left( p:1-{{m}_{2,n}} \right)  \\
   -  \\
\end{matrix} \right]
\end{equation}
\hrulefill
\end{figure*}

An analytical solution for the distribution of the sum of Double GG RVs is a very cumbersome task if not impossible. Thus, in this paper, we derive a novel upper bound for the distribution of the sum of Double GG distributed RVs. Particularly, a useful expression for the distribution of the product of Double GG distributed RVs is first obtained. Then, based on a well-known inequality between arithmetic and geometric means, a closed-form union upper bound for the distribution of the sum of Double GG distributed RVs is derived. This derived distribution is used to study the error rate performance of FSO systems with MISO/SIMO with EGC receivers employing intensity modulation/direct detection (IM/DD) with on-off keying (OOK) over independent and not necessarily identically distributed (i.n.i.d.) Double GG turbulence channels.

The rest of the paper is organized as follows: In Section II, the statistical characteristics of the Double GG distribution are provided, and an upper bound on the pdf of the sum of Double GG variates is presented. In Section III, we introduce the SIMO FSO system model. In Section IV, the BER expressions for SIMO FSO links are provided. In Section V, we present numerical results to confirm the accuracy of the derived expressions and demonstrate the advantages of employing spatial diversity over SISO links. Finally, Section VI concludes the paper.
\section{Definition and Statistical Characteristics}
\subsection{Double GG distribution}
Let $\left\{ {{I}_{n}} \right\}_{n=1}^{N}$ be $N$ statistically independent but not necessarily identically distributed Double GG RVs whose probability density function (pdf) follows (\ref{eq1}) at the top of the page [\citen{kashani}, Eq. (5)].

In (\ref{eq1}), $G\left[ . \right]$ is the Meijer’s G-function\footnote{Meijer’s G-function is a standard built-in function in mathematical software packages such as MATLAB, MAPLE and MATHEMATICA. If required, this function can be also expressed in terms of the generalized hypergeometric functions using [\citen{24}, Eqs.(9.303-304)].}  [\citen{24}, Eq.(9.301)], $\Delta \left( j;x \right)$ is defined as $\Delta \left( j;x \right)\triangleq x/j\ ,...,\left( x+j-1 \right)/j\ $, ${{p}_{n}}$ and ${{q}_{n}}$ are positive integer numbers that satisfy ${{p}_{n}}/{{q}_{n}}\ ={{\gamma }_{1,n}}/{{\gamma }_{2,n}}$, and ${{m}_{i,n}}\ge 0.5$, $i=1,2$, is a distribution shaping parameter. The distribution parameters ${{\gamma }_{i,n}}$ and ${{\Omega }_{i,n}}$ of the Double GG model can be identified using the following equations
\begin{align}\label{eq2}
&{{\Omega }_{i,n}}={{\left( \frac{\Gamma \left( {{m}_{i,n}} \right)}{\Gamma \left( {{m}_{i,n}}+1/{{\gamma }_{i,n}} \right)} \right)}^{{{\gamma }_{i,n}}}}{{m }_{i,n}},~~i=1,2\\\label{eq3a}
\newsubeqblock
\mysubeq &\sigma _{x,n}^{2}=\frac{\Gamma \left( {{m }_{1,n}}+{2}/{{{\gamma }_{1,n}}}\; \right)\Gamma \left( {{m }_{1,n}} \right)}{{{\Gamma }^{2}}\left( {{m }_{1,n}}+{1}/{{{\gamma }_{1,n}}}\; \right)}-1\\\label{eq3b}
\mysubeq &\sigma _{y,n}^{2}=\frac{\Gamma \left( {{m }_{2,n}}+{2}/{{{\gamma }_{2,n}}}\; \right)\Gamma \left( {{m }_{2,n}} \right)}{{{\Gamma }^{2}}\left( {{m }_{2,n}}+{1}/{{{\gamma }_{2,n}}}\; \right)}-1
\end{align}
where $\sigma _{x,n}^{2}$ and $\sigma _{y,n}^{2}$ are normalized variances of small and large scale irradiance fluctuations, respectively. The Double GG distribution accurately describes irradiance fluctuations over atmospheric channels under a wide range of turbulence conditions (weak to strong) \cite{kashani}. Furthermore, it is very generic and includes most commonly used fading models proposed in the literature as special cases such as Gamma-Gamma (${{\gamma }_{i,n}}=1$, ${{\Omega }_{i,n}}=1$), Double-Weibull (${{m}_{i,n}}=1$), and K channel (${{\gamma }_{i,n}}=1$, ${{\Omega }_{i,n}}=1$, ${{m}_{2,n}}=1$). In addition, for the limiting case of ${{\gamma }_{i,n}}\to 0$ and ${{m}_{i,n}}\to \infty $, Double GG pdf coincides with the log-normal pdf.

The cumulative distribution function (cdf) of Double GG distribution can be derived from (\ref{eq1}) as in (\ref{eq4}) at the top of the next page \cite{kashani}.
\begin{figure*}[t]
\begin{equation}\label{eq4}
{{F}_{{{I}_{n}}}}\left( {{I}_{n}} \right)=\frac{p_{n}^{{{m}_{2}}-1/2}q_{n}^{{{m}_{1}}-1/2}{{\left( 2\pi  \right)}^{{1-\left( {{p}_{n}}+{{q}_{n}} \right)}/{2}\;}}}{\Gamma \left( {{m}_{1,n}} \right)\Gamma \left( {{m}_{2,n}} \right)}G_{1,{{p}_{n}}+{{q}_{n}}+1}^{{{p}_{n}}+{{q}_{n}},1}\left[ {{\left( \frac{I_{n}^{{{\gamma }_{2,n}}}}{{{\Omega }_{2,n}}} \right)}^{{{p}_{n}}}}\frac{m_{1,n}^{{{q}_{n}}}m_{2,n}^{{{p}_{n}}}}{p_{n}^{{{p}_{n}}}q_{n}^{{{q}_{n}}}\Omega _{1,n}^{{{q}_{n}}}}|\begin{matrix}
   1  \\
   \Delta \left( {{q}_{n}}:{{m}_{1,n}} \right),\Delta \left( {{p}_{n}}:{{m}_{2,n}} \right),0  \\
\end{matrix} \right]
\end{equation}
\hrulefill
\end{figure*}
\subsection{An Upper-Bound for the Distribution of the Sum of Double GG Variates}
Let us define a new R.V $R$, as the product of $N$ Double GG R.Vs ${{I}_{n}}$, i.e.,
\begin{equation}\label{eq5}
R\triangleq \prod\limits_{n=1}^{N}{{{I}_{n}}}.
\end{equation}

The Double GG distribution considers irradiance fluctuations as the product of small-scale and large-scale fluctuations which are both governed by Generalized Gamma (GG) distributions, i.e. ${{I}_{n}}={{U}_{x,n}}{{U}_{y,n}}$, where ${{U}_{x,n}}$ and ${{U}_{y,n}}$ are statistically independent arising respectively from large-scale and small scale turbulent eddies. Thus, $R$ can be expressed as the product of $2N$ GG R.Vs ${{U}_{n}}$, i.e.,
\begin{equation}\label{eq6}
R=\prod\limits_{n=1}^{2N}{{{U}_{n}}}.
\end{equation}
The pdf of $R$ can be obtained using the statistical model proposed in \cite{25} as
\begin{equation}\label{eq7}
{{f}_{R}}\left( r \right)=\frac{\alpha \xi }{r}G_{0,\beta }^{\beta ,0}\left[ \frac{{{r}^{\alpha }}}{\omega }\left| \begin{matrix}
   -  \\
   {{J}_{\alpha }}\left( {{\gamma }_{1:2N}},{{m}_{1:2N}} \right)  \\
\end{matrix} \right. \right]
\end{equation}
where $\xi $, $\omega$ and ${{J}_{\alpha }}\left( {{\gamma }_{1:2N}},{{m}_{1:2N}} \right)$ are defined as
\begin{align}\label{eq8}
&\xi ={{\left( \sqrt{2\pi } \right)}^{2N-\beta }}\prod\limits_{l=1}^{2N}{\frac{{{\left( {\alpha }/{{{\gamma }_{l}}}\; \right)}^{{{m}_{l}}-1/2}}}{\Gamma \left( {{m}_{l}} \right)}}\\\label{eq9}
&\omega ={{\prod\limits_{l=1}^{2N}{\left( \frac{\alpha {{\Omega }_{l}}}{{{m}_{l}}{{\gamma }_{l}}} \right)}}^{{\alpha }/{{{\gamma }_{l}}}\;}}\\\label{eq10}
&{{J}_{\alpha }}\left( {{\gamma }_{1:2N}},{{m}_{1:2N}} \right)\triangleq\\\nonumber
&\Delta \left( {}^{\alpha }/{}_{{{\gamma }_{1}}};{{m}_{1}} \right),\Delta \left( {}^{\alpha }/{}_{{{\gamma }_{2}}};{{m}_{2}} \right),\ldots ,\Delta \left( {}^{\alpha }/{}_{{{\gamma }_{2N}}};{{m}_{2N}} \right)
\end{align}
In (\ref{eq7}), $\alpha $ and $\beta $ are two positive integers defined as
\begin{align}\label{eq11}
&\alpha \triangleq \prod\limits_{l=1}^{2N}{{{k}_{l}}}\\\label{eq12}
&\beta \triangleq \alpha \sum\limits_{l=1}^{2N}{\frac{1}{{{\gamma }_{l}}}}
\end{align}
under the constraint that
\begin{equation}\label{eq13}
{{l}_{l}}=\frac{1}{{{\gamma }_{l}}}\prod\limits_{i=1}^{l}{{{k}_{i}}}
\end{equation}
is a positive integer with ${{k}_{i}}$ being also a positive integer.

The cdf of $R$ can be derived from (\ref{eq7}) as
\begin{equation}\label{eq14}
{{F}_{R}}\left( r \right)=\xi G_{1,\beta +1}^{\beta ,1}\left[ \frac{{{r}^{\alpha }}}{\omega }\left| \begin{matrix}
   1  \\
   {{J}_{\alpha }}\left( {{\gamma }_{1:2N}},{{m}_{1:2N}} \right),0  \\
\end{matrix} \right. \right].
\end{equation}

Using the well-known inequality between arithmetic and geometric means, i.e. ${{A}_{N}}\ge {{G}_{N}}$, with
\begin{equation}\label{eq15}
{{A}_{N}}=\frac{1}{N}\sum\limits_{n=1}^{n}{{{I}_{n}}}
\end{equation}
and
\begin{equation}\label{eq16}
{{G}_{N}}=\prod\limits_{n=1}^{N}{I_{n}^{{1}/{N}\;}}
\end{equation}
a lower-bound for R.V $Z$ deﬁned as the sum of Double GG R.Vs, i.e.,
\begin{equation}\label{eq17}
Z\triangleq \sum\limits_{n=1}^{N}{{{I}_{n}}}
\end{equation}
can be obtained as
\begin{equation}\label{eq18}
Z\ge N{{R}^{{1}/{N}}}.
\end{equation}
Considering Eqs. (\ref{eq14}) and (\ref{eq18}), the cdf of $Z$ is upper bounded as
\begin{equation}\label{eq19}
{{F}_{Z}}\left( r \right)\le \xi G_{1,\beta +1}^{\beta ,1}\left[ \frac{{{\left( {r}/{N}\; \right)}^{\alpha N}}}{\omega }\left| \begin{matrix}
   1  \\
   {{J}_{\alpha }}\left( {{\gamma }_{1:2N}},{{m}_{1:2N}} \right),0  \\
\end{matrix} \right. \right].
\end{equation}
By taking the first derivative of (\ref{eq19}) with respect to $r$, an upper bound for the pdf of $Z$ can be obtained in closed-form as
\begin{equation}\label{eq20}
{{f}_{Z}}\left( r \right)\le f_{Z}^{*}\left( r \right)
\end{equation}
where $f_{Z}^{*}\left( r \right)$ is defined as
\begin{equation}\label{eq21}
f_{Z}^{*}\left( r \right)=\frac{N\alpha \xi }{r}G_{0,\beta }^{\beta ,0}\left[ \frac{{{\left( {r}/{N}\; \right)}^{\alpha N}}}{\omega }\left| \begin{matrix}
   -  \\
   {{J}_{\alpha }}\left( {{\gamma }_{1:2N}},{{m}_{1:2N}} \right)  \\
\end{matrix} \right. \right].
\end{equation}
\section{System Model}
We consider an FSO system employing IM/DD with OOK where the information signal is transmitted via one aperture and received by $N$ apertures (i.e., SIMO) over the Double GG channel. We assume EGC receivers where the receiver adds the receiver branches. The received signal is then given by
\begin{equation}\label{eq22}
r=\eta x\sum\limits_{n=1}^{N}{{{I}_{n}}}+{{\upsilon }_{n}},\,\,\,n=1,\ldots ,N
\end{equation}
where $x$ represents the information bits and can either be 0 or 1, ${{\upsilon }_{n}}$ is the Additive White Gaussian noise (AWGN) at the ${{n}^{\text{th}}}$ receive aperture with zero mean and variance $\sigma _{\upsilon}^{2}={{N}_{0}}/2$ , and $\eta$ is the optical-to-electrical conversion coefficient. Here, ${{I}_{n}}$ is the normalized irradiance from the transmitter to the ${{n}^{\text{th}}}$ receive aperture whose pdf follows (\ref{eq1}). We should emphasize that the performance of SIMO under the assumption of equal gain combining is equivalent to that of MISO FSO links.
\section{BER Performance}
\subsection{Upper bound expression}
The optimum decision metric for OOK is given by \cite{21}
\begin{equation}\label{eq23}
P\left( r|\text{on,}{{\text{I}}_{n}} \right)\underset{\text{off}}{\mathop{\overset{\text{on}}{\mathop{\lessgtr }}\,}}\,P\left( r|\text{off,}{{\text{I}}_{n}} \right)
\end{equation}
where $r$ is the received signal vector. Following the same approach as \cite{20,21}, the conditional bit error probabilities are given by (see \cite{20} for details of derivation)
\begin{equation}\label{eq24}
{{P}_{e,\text{MISO}}}(\text{off }|{{I}_{n}})={{P}_{e,\text{MISO}}}(\text{on }|{{I}_{n}})=\frac{1}{2}\operatorname{erfc}\left( \frac{\sqrt{{\bar{\gamma }}}}{2N}\sum\limits_{n=1}^{N}{{{I}_{n}}} \right)
\end{equation}
where $\bar{\gamma }$ is the average electrical SNR obtained as $\bar{\gamma }={{{\eta }^{2}}}/{{{N}_{0}}}$. Therefore, the average error rate can be expressed as
\begin{equation}\label{eq25}
{{P}_{\text{SIMO,ECG}}}=\frac{1}{2}\int\limits_{\mathbf{I}}{{{f}_{\mathbf{I}}}}\left( \mathbf{I} \right)\operatorname{erfc}\left( \frac{\sqrt{{\bar{\gamma }}}}{2N}\sum\limits_{n=1}^{N}{{{I}_{n}}} \right)d\mathbf{I}
\end{equation}
where ${{f}_{\mathbf{I}}}\left( \mathbf{I} \right)$ is the joint pdf of vector $\mathbf{I}=\left( {{I}_{1}},{{I}_{2}},\ldots ,{{I}_{N}} \right)$. The factor $N$ is used to ensure that the sum of the $N$ receive aperture areas is the same as the area of the receive aperture of the SISO link for a fair comparison. The integral expressed in (\ref{eq25}) does not yield a closed-form solution even for simpler turbulence distributions. However, an upper bound on (\ref{eq25}) can be obtained by considering (\ref{eq20}) and (\ref{eq21}) as
\begin{equation}\label{eq26}
{{P}_{\text{MISO}}}\le \frac{1}{2}\int\limits_{0}^{\infty }{f_{Z}^{*}\left( z \right)}\operatorname{erfc}\left( \frac{\sqrt{{\bar{\gamma }}}z}{2N} \right)dz.
\end{equation}
The above integral can be evaluated in closed form by first expressing the $\operatorname{erfc}\left( . \right)$ in terms of the Meijer G-function presented in [\citen{29}, eq. (11)] as
\begin{equation}\label{eq27}
\operatorname{erfc}\left( \sqrt{x} \right)=\frac{1}{\sqrt{\pi }}G_{1,2}^{2,0}\left[ x\left| \begin{matrix}
   1  \\
   0,{1}/{2}\;  \\
\end{matrix} \right. \right].
\end{equation}
Then, a closed-form expression for (\ref{eq26}) is obtained using [\citen{29}, Eq. (21)] as
\begin{align}\label{eq28}
&{{P}_{\text{SIMO,ECG}}}\le \frac{N\alpha \xi {{q}^{\mu }}}{2\sqrt{2}s{{\left( \sqrt{2\pi } \right)}^{s+(q-1)\beta }}}\\\nonumber
&\times G_{2s,q\beta +s}^{q\beta ,2s}\left[ \frac{{{\left( 4{{{\bar{\gamma }}}^{-1}}s{{N}^{2}} \right)}^{s}}}{{{\left( \omega {{q}^{\beta }}{{N}^{\alpha N}} \right)}^{q}}}\left| \begin{matrix}
   \Delta \left( s,1 \right),\Delta \left( s,{1}/{2}\; \right)  \\
   {{K }_{q}}\left( {\alpha }/{{{\gamma }_{1:2N}}}\;,{{m}_{1:2N}} \right),\Delta \left( s,0 \right)  \\
\end{matrix} \right. \right]
\end{align}
where $s$ and $q$ are positive integer numbers that satisfy ${{s}/{q}\;=\alpha N}/{2}$, and $\mu $ and ${{K}_{q}}\left( {\alpha }/{{{\gamma }_{1:2N}}}\;,{{m}_{1:2N}} \right)$ are defined as
\begin{align}\label{eq29}
&{{K}_{q}}\left( {\alpha }/{{{\gamma }_{1:2N}}}\;,{{m}_{1:2N}} \right)=\\\nonumber
&\begin{cases}
  & {{J}_{\alpha }}\left( {{\gamma }_{1:2N}},{{m}_{1:2N}} \right)\,\,\,\,\,\,\,\,\,\,\,\,\,\,\,\,\,\,\,\,\,\,\,\,\,\,\,\,\,\,\,\,\,\,\,\,\,\,\,\,\,\,\,\,\,\,\,\,\,q=1 \\
 & \left\{ \frac{{{J}_{\alpha }}\left( {{\gamma }_{1:2N}},{{m}_{1:2N}} \right)}{2},\frac{{{J}_{\alpha }}\left( {{\gamma }_{1:2N}},{{m}_{1:2N}} \right)-1}{2} \right\}\,\,q=2 \\
\end{cases}
\end{align}
\begin{equation}\label{eq30}
\mu =\sum\limits_{l=1}^{2N}{{{m}_{l}}}-N+1.
\end{equation}
The derived upper-bounded BER expression in (\ref{eq28}) for SIMO FSO systems with ECG can be seen as a generalization of BER results over other atmospheric turbulence models. Specially, if we insert ${{\gamma }_{i}}=1$ and ${{\Omega }_{i}}=1$ in (\ref{eq28}), we obtain an upper bound on BER expression over Gamma-Gamma channel. Setting ${{m}_{i}}=1$ in (\ref{eq28}), we obtain an upper bound for BER for Double Weibull channel. Similarly, for ${{\gamma }_{i}}=1$, ${{\Omega }_{i}}=1$ and ${{m}_{2i}}=1$, an upper-bounded BER expression for K-channel is obtained.
\subsection{Diversity Gain and Asymptotic analysis}
Although Meijer’s G-function can be expressed in terms of more tractable generalized hypergeometric functions, (\ref{eq28}) appears to be complex and the impact of the basic system and channel parameters on performance is not very clear. However, for large SNR values, the asymptotic behavior of the system performance is dominated by the behavior of the pdf near the origin, i.e. $f_{Z}^{*}\left( z \right)$ at $z\to 0$ \cite{garcia}. Thus, employing a series expansion corresponding to the Meijer’s G-function [\citen{wol}, Eq. (07.34.06.0006.01)], $f_{Z}^{*}\left( z \right)$ given in (\ref{eq21}) can be approximated by a single polynomial term as
\begin{align}\nonumber
&f_{Z}^{*}\left( z \right)\approx \frac{N\alpha \xi }{{{\left( \omega {{N}^{\alpha N}} \right)}^{\min \left\{ \frac{{{m}_{1}}{{\gamma }_{1}}}{\alpha },\cdots ,\frac{{{m}_{2N}}{{\gamma }_{2N}}}{\alpha } \right\}}}}\\\label{eq31}
&\prod\limits_{\begin{smallmatrix}
 j=1 \\
 j\ne k
\end{smallmatrix}}^{\beta }{\Gamma \left( {{c}_{j}}-{{c}_{k}} \right){{z}^{N\min \left\{ {{m}_{1}}{{\gamma }_{1}},\cdots ,{{m}_{2N}}{{\gamma }_{2N}} \right\}-1}}}
\end{align}
where ${{c}_{k}}$ and ${{c}_{j}}$ are defined as
\begin{equation}\label{eq32}
{{c}_{k}}=\min \left\{ \frac{{{m}_{1}}{{\gamma }_{1}}}{\alpha },\cdots ,\frac{{{m}_{2N}}{{\gamma }_{2N}}}{\alpha } \right\}
\end{equation}
\begin{align}\nonumber
&{{c}_{j}}\in\left\{ \Delta \left( {}^{\alpha }/{}_{{{\gamma }_{2}}};{{m}_{2}} \right),\ldots ,\right.\\\label{eq33}
&\left.\Delta \left( {}^{\alpha }/{}_{{{\gamma }_{2N}}};{{m}_{2N}} \right) \right\}\backslash \min \left\{ \frac{{{m}_{1}}{{\gamma }_{1}}}{\alpha },\cdots ,\frac{{{m}_{2N}}{{\gamma }_{2N}}}{\alpha } \right\}.
\end{align}
It must be noted that (\ref{eq31}) is only valid for independent and not identically distributed Double GG turbulence channels. Based on Eqs. (\ref{eq26}) and (\ref{eq31}) and at high SNRs, the average BER can be well approximated as
\begin{equation}\label{eq34}
{{P}_{\text{SIMO,ECG}}}\approx \frac{\Gamma \left( \left( 1+N\alpha {{c}_{k}} \right)/2 \right)\xi }{2\sqrt{\pi }{{c}_{k}}{{\left( \omega {{N}^{\alpha N}} \right)}^{{{c}_{k}}}}}\prod\limits_{\begin{smallmatrix}
 j=1 \\
 j\ne k
\end{smallmatrix}}^{\beta }{\Gamma \left( {{c}_{j}}-{{c}_{k}} \right)}{{\left( \frac{2N}{\sqrt{{\bar{\gamma }}}} \right)}^{N\alpha {{c}_{k}}}}
\end{equation}
Therefore, the diversity order of FSO links with $N$ receive apertures employing equal gain combining is obtained as $0.5N\min \left\{ {{m}_{1}}{{\gamma }_{1}}\cdots ,{{m}_{2N}}{{\gamma }_{2N}} \right\}$.
\begin{figure}[t]
\centering
\includegraphics[width = 8cm, height = 7.5cm]{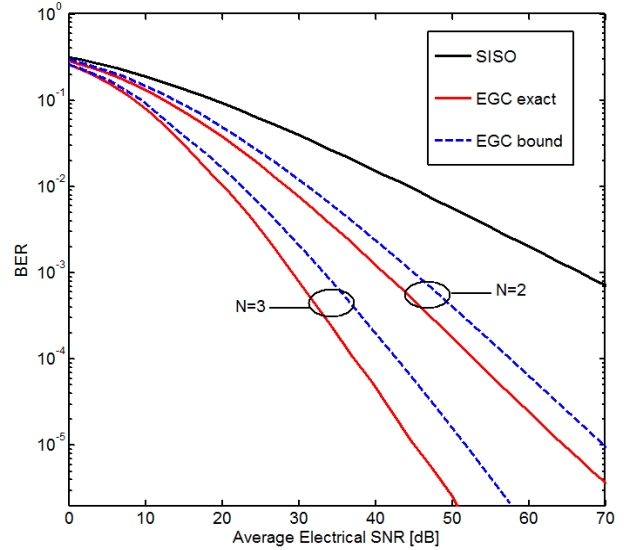}
\caption{ Average BER of EGC and SISO for plane wave assuming i.i.d. turbulent channel defined as channel $b$.}
\end{figure}
\begin{figure}[t]
\centering
\includegraphics[width = 8cm, height = 7.5cm]{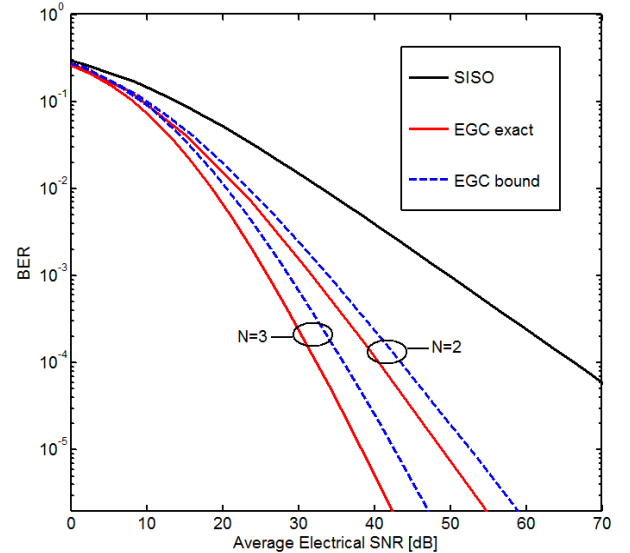}
\caption{ Average BER of EGC and SISO for spherical wave assuming i.i.d. turbulent channel defined as channel $c$.}
\end{figure}
\section{Numerical Results}
In this section, we present analytical and Monte-Carlo simulation results using the previous mathematical analysis for the performance of SIMO FSO links employing EGC receivers over Double GG channels. The performance improvements over SISO systems are further quantified. Similar as in \cite{ICC}, we consider the following four scenarios of atmospheric turbulence conditions reported in \cite{kashani}
\begin{itemize}
   \item  \textbf{Channel a}: \emph{Plane wave and moderate irradiance fluctuations} with ${{\gamma }_{1}}=2.1690$, ${{\gamma }_{2}}=0.8530$, ${{m}_{1}}=0.55$, ${{m}_{2}}=2.35$, ${{\Omega }_{1}}=1.5793$, ${{\Omega }_{2}}=0.9671$, $p=28$ and $q=11$
  \item  \textbf{Channel b}: \emph{Plane wave and strong irradiance fluctuations} with ${{\gamma }_{1}}=1.8621$, ${{\gamma }_{2}}=0.7638$, ${{m}_{1}}=0.5$, ${{m}_{2}}=1.8$, ${{\Omega }_{1}}=1.5074$, ${{\Omega }_{2}}=0.9280$, $p=17$ and $q=7$.
  \item  \textbf{Channel c}: \emph{Spherical wave and moderate irradiance fluctuations} with ${{\gamma }_{1}}=0.9135$, ${{\gamma }_{2}}=1.4385$, ${{m}_{1}}=2.65$, ${{m}_{2}}=0.85$, ${{\Omega }_{1}}=0.9836$ and ${{\Omega }_{2}}=1.1745$, $p=7$ and $q=11$.
  \item \textbf{Channel d}: \emph{Spherical wave and strong irradiance fluctuations} with ${{\gamma }_{1}}=0.4205$, ${{\gamma }_{2}}=0.6643$, ${{m}_{1}}=3.2$, ${{m}_{2}}=2.8$, ${\Omega_{1}}=0.8336$ and ${{\Omega }_{2}}=0.9224$, $p=7$ and $q=11$.
\end{itemize}

Figs. 1-2 demonstrate the average BER over i.i.d. channels defined as channel $b$ and channel $c$, respectively. In order to verify the tightness of the bound, we present upper bound analytical results obtained through (\ref{eq28}) along with the Monte-Carlo simulation of (\ref{eq25}). As a benchmark, the average BER of SISO FSO link obtained through [\citen{kashani}, Eq. 24] is also included in these figures. As clearly seen from Figs. 1-2, the numerical results for the bounds are close to the equivalent simulated ones which represent the exact BER. For instance, at a target bit error rate of ${{10}^{-5}}$, gaps between the exact and the upper bound curves are 5.2 dB and 6.6 dB respectively for $N=2$ and $3$ receive apertures employing EGC for channel $b$. Similarly, for channel $c$, at a BER of ${{10}^{-5}}$, the gaps are respectively 3.8 dB and 4.3 dB for $N=2$ and $3$ receive apertures employing EGC. This observation clearly demonstrates the accuracy of the proposed bound. It is also illustrated that the upper bound becomes tighter as the number of receive apertures decreases. This is expected as the upper bound and the exact BER curves coincide for $N = 1$. In addition, we observe that multiple receive apertures deployment employing EGC signiﬁcantly improves the performance. Specially, at a target bit error rate of ${{10}^{-5}}$, we observe performance improvements of 46.8 dB and 66.8 dB for SIMO FSO links with $N=2$ and $3$ receive apertures with respect to the SISO transmission over channel $b$. Similarly, for channel $c$, at a BER of ${{10}^{-5}}$, impressive performance improvements of 51.1 dB and 63.9 dB are achieved for SIMO links with $N=2$ and 3 employing EGC compared to the SISO deployment.
\begin{figure}
\centering
\includegraphics[width = 8cm, height = 7.5cm]{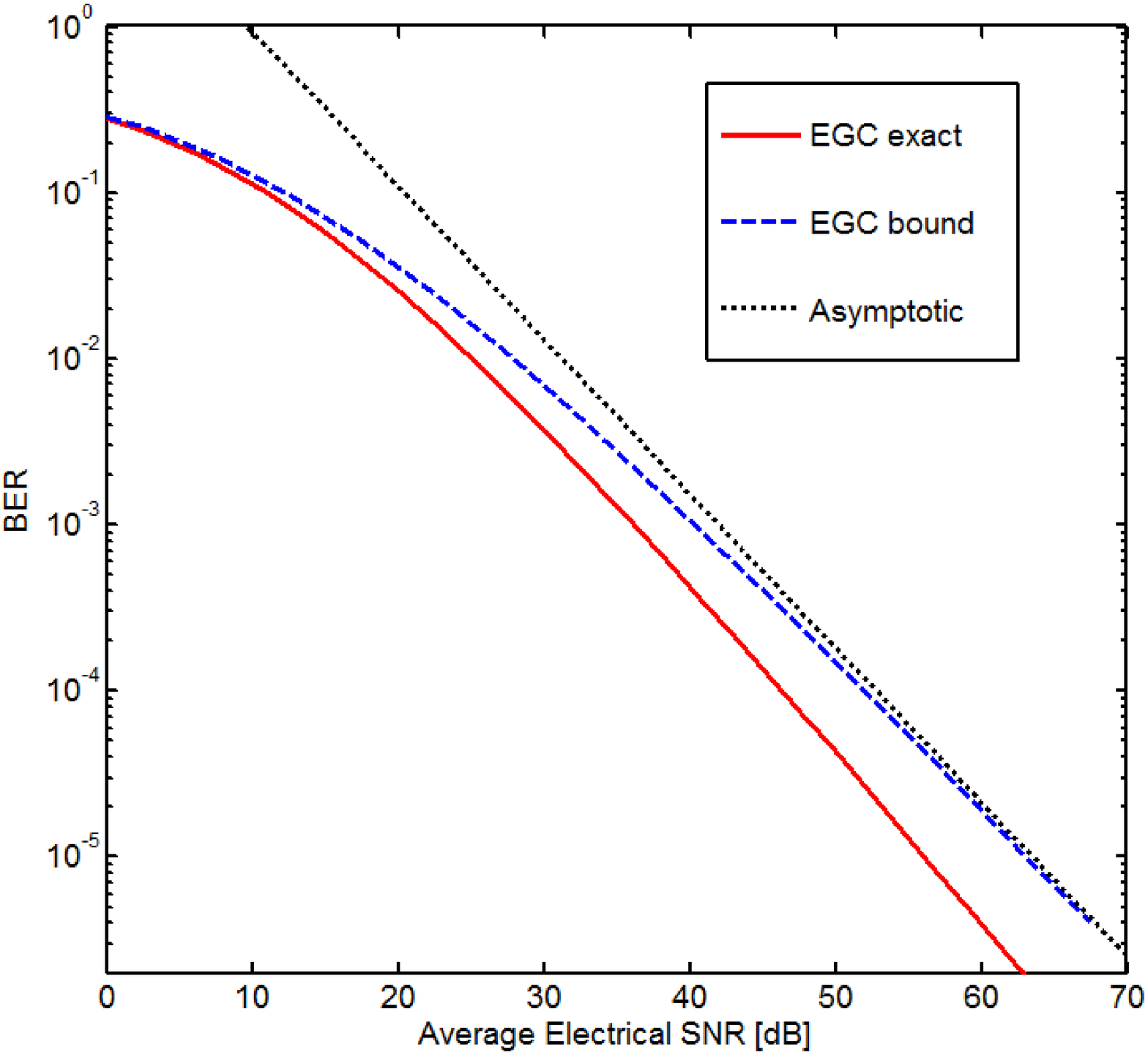}
\caption{Average BER of EGC over two i.n.i.d. atmospheric turbulence channels defined as channel $a$ and channel $b$.}
\end{figure}
\begin{figure}
\centering
\includegraphics[width = 8cm, height = 7.5cm]{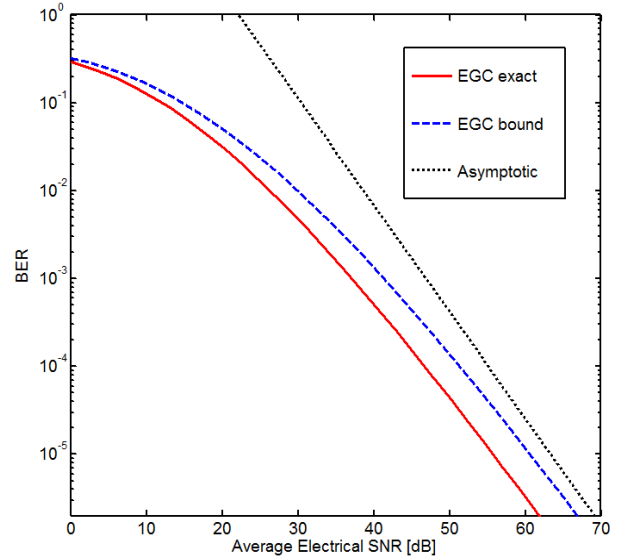}
\caption{Average BER of EGC over two i.n.i.d. atmospheric turbulence channels defined as channel $c$ and channel $d$.}
\end{figure}

Figs. 3-4 illustrate the BER performance of SIMO FSO links employing EGC receivers over non-identically distributed (i.n.i.d.) Double GG channels. Similar to i.i.d. results, our upper bound closed-form expression yields a close match to simulation results. For example, at a BER of ${{10}^{-5}}$ in SIMO links with $N=2$ over i.n.i.d. channels $a$ and $b$, the difference between the exact and the upper bound curves is 7 dB. For i.n.i.d. channels $c$ and $d$ and at a BER of ${{10}^{-5}}$, this gap is 5.8 dB with $N=2$. We also compare the performance of i.n.i.d. case with respect to i.i.d. case presented in Figs. 1-2. As an example, to achieve a BER of ${{10}^{-5}}$ in SIMO links over i.n.i.d. channels $a$ and $b$, we need 8.5 dB less in comparison to i.i.d. case as channel $a$ is less severe than channel $b$. Note that in Fig. 1, we assume that both of the two channels between transmitter and receivers are described by channel $b$. On the other hand, to achieve a BER of ${{10}^{-5}}$ for SIMO links with $N=2$ over i.n.i.d. channels $c$ and $d$, we need 6.8 dB more in comparison to i.i.d channels as the channel $d$ exhibits harsher conditions than channel $c$. Note that in Fig. 2, both of the channels between the transmitter and receivers are described by channel $c$. It can be further observed that asymptotic bounds on the BER become tighter at high enough SNR values confirming the accuracy and usefulness of the asymptotic expression given in (\ref{eq34}).
\section{Conclusions}\label{Con}
In this paper, we have derived a closed-form union upper bound for the pdf of the sum of Double GG distributed RVs. Using this bound, we have investigated the BER performance of FSO links with receive diversity employing equal gain combining over Double GG turbulence channels. An efficient and unified upper bound for the average BER of SIMO FSO systems with EGC receiver has been obtained which generalizes BER results over other atmospheric turbulence models as special cases. Based on the asymptotical performance analysis, we have further derived diversity gains for SIMO FSO systems under consideration. We have presented BER performance based on analytical and numerical simulation results. Further comparisons between numerical and analytical results have confirmed the accuracy and usefulness of the derived results.
\balance
\bibliographystyle{IEEEtran}
\bibliography{ref}

\begin{thebibliography}{10}
\providecommand{\url}[1]{#1}
\csname url@samestyle\endcsname
\providecommand{\newblock}{\relax}
\providecommand{\bibinfo}[2]{#2}
\providecommand{\BIBentrySTDinterwordspacing}{\spaceskip=0pt\relax}
\providecommand{\BIBentryALTinterwordstretchfactor}{4}
\providecommand{\BIBentryALTinterwordspacing}{\spaceskip=\fontdimen2\font plus
\BIBentryALTinterwordstretchfactor\fontdimen3\font minus
  \fontdimen4\font\relax}
\providecommand{\BIBforeignlanguage}[2]{{%
\expandafter\ifx\csname l@#1\endcsname\relax
\typeout{** WARNING: IEEEtran.bst: No hyphenation pattern has been}%
\typeout{** loaded for the language `#1'. Using the pattern for}%
\typeout{** the default language instead.}%
\else
\language=\csname l@#1\endcsname
\fi
#2}}
\providecommand{\BIBdecl}{\relax}
\BIBdecl

\bibitem{1}
Z.~Hajjarian, J.~Fadlullah, and M.~Kavehrad, ``Mimo free space optical
  communications in turbid and turbulent atmosphere,'' \emph{Journal of
  Communications}, vol.~4, no.~8, pp. 524--532, 2009.

\bibitem{2}
S.~Arnon, J.~R. Barry, G.~K. Karagiannidis, R.~Schober, and M.~Uysal~(Eds.),
  \emph{Advanced Optical Wireless Communication}.\hskip 1em plus 0.5em minus
  0.4em\relax Cambridge university press, 2012.

\bibitem{kashani2}
M.~A. Kashani and M.~Uysal, ``Outage performance of fso multi-hop parallel
  relaying,'' in \emph{IEEE Signal Processing and Communications Applications
  Conference (SIU)}.\hskip 1em plus 0.5em minus 0.4em\relax IEEE, 2012, pp.
  1--4.

\bibitem{3}
L.~C. Andrews and R.~L. Phillips, \emph{Laser beam propagation through random
  media}.\hskip 1em plus 0.5em minus 0.4em\relax Society of Photo Optical,
  2005, vol. 152.

\bibitem{4}
M.~A. Kashani, M.~Safari, and M.~Uysal, ``Optimal relay placement and diversity
  analysis of relay-assisted free-space optical communication systems,''
  \emph{IEEE/OSA Journal of Optical Communications and Networking}, vol.~5,
  no.~1, pp. 37--47, 2013.

\bibitem{5}
M.~A. Kashani and M.~Uysal, ``Outage performance and diversity gain analysis of
  free-space optical multi-hop parallel relaying,'' \emph{IEEE/OSA Journal of
  Optical Communications and Networking}, vol.~5, no.~8, pp. 901--909, 2013.

\bibitem{6}
M.~Karimi and M.~Nasiri-Kenari, ``Ber analysis of cooperative systems in
  free-space optical networks,'' \emph{Journal of Lightwave Technology},
  vol.~27, no.~24, pp. 5639--5647, 2009.

\bibitem{6.1}
M.~A. Kashani, M.~M. Rad, M.~Safari, and M.~Uysal, ``All-optical
  amplify-and-forward relaying system for atmospheric channels,'' \emph{IEEE
  Communications Letters}, vol.~16, no.~10, pp. 1684--1687, 2012.

\bibitem{6.2}
C.~Abou-Rjeily, ``Performance analysis of selective relaying in cooperative
  free-space optical systems,'' \emph{IEEE Journal of Lightwave Technology},
  vol.~31, no.~18, pp. 2965--2973, 2013.

\bibitem{kashani3}
M.~A. Kashani, M.~Safari, and M.~Uysal, ``Optimal relay placement in
  cooperative free-space optical communication systems.'' in \emph{IEEE WCNC},
  2012, pp. 995--999.

\bibitem{7}
E.~Jakeman and P.~Pusey, ``Significance of k distributions in scattering
  experiments,'' \emph{Physical Review Letters}, vol.~40, no.~9, pp. 546--550,
  1978.

\bibitem{8}
L.~Andrews and R.~Phillips, ``Mathematical genesis of the ik distribution for
  random optical fields,'' \emph{Journal of Optical Society of America A},
  vol.~3, no.~11, pp. 1912--1919, 1986.

\bibitem{9}
J.~H. Churnside and S.~F. Clifford, ``Log-normal rician probability-density
  function of optical scintillations in the turbulent atmosphere,''
  \emph{Journal of Optical Society of America A}, vol.~4, no.~10, pp.
  1923--1930, 1987.

\bibitem{10}
M.~Al-Habash, L.~C. Andrews, and R.~L. Phillips, ``Mathematical model for the
  irradiance probability density function of a laser beam propagating through
  turbulent media,'' \emph{Optical Engineering}, vol.~40, no.~8, pp.
  1554--1562, 2001.

\bibitem{11}
A.~Jurado-Navas, J.~M. Garrido-Balsells, J.~F. Paris, and A.~Puerta-Notario,
  ``A unifying statistical model for atmospheric optical scintillation,''
  \emph{arXiv preprint arXiv:1102.1915}, 2011.

\bibitem{12}
N.~D. Chatzidiamantis, H.~G. Sandalidis, G.~K. Karagiannidis, S.~A.
  Kotsopoulos, and M.~Matthaiou, ``New results on turbulence modeling for
  free-space optical systems,'' in \emph{IEEE International Conference on
  Telecommunications (ICT)}, 2010, pp. 487--492.

\bibitem{kashani}
M.~Amini~Kashani, M.~Uysal, and M.~Kavehrad, ``A novel statistical channel
  model for turbulence-induced fading in free-space optical systems,''
  \emph{Journal of Lightwave Technology}, vol.~PP, no.~99, pp. 1--1, 2015.

\bibitem{15}
E.~Bayaki, R.~Schober, and R.~K. Mallik, ``Performance analysis of mimo
  free-space optical systems in gamma-gamma fading,'' \emph{IEEE Transactions
  on Communications}, vol.~57, no.~11, pp. 3415--3424, 2009.

\bibitem{20}
T.~A. Tsiftsis, H.~G. Sandalidis, G.~K. Karagiannidis, and M.~Uysal, ``Optical
  wireless links with spatial diversity over strong atmospheric turbulence
  channels,'' \emph{IEEE Transactions on Wireless Communications}, vol.~8,
  no.~2, pp. 951--957, 2009.

\bibitem{21}
S.~M. Navidpour, M.~Uysal, and M.~Kavehrad, ``Ber performance of free-space
  optical transmission with spatial diversity,'' \emph{IEEE Transactions on
  Wireless Communications}, vol.~6, no.~8, pp. 2813--2819, 2007.

\bibitem{Farid}
A.~A. Farid and S.~Hranilovic, ``Diversity gain and outage probability for mimo
  free-space optical links with misalignment,'' \emph{IEEE Transactions on
  Communications}, vol.~60, no.~2, pp. 479--487, 2012.

\bibitem{Antonio}
A.~Garc{\'y}a-Zambrana, C.~Castillo-V{\ss}zquez, and B.~Castillo-V{\ss}zquez,
  ``Space-time trellis coding with transmit laser selection for fso links over
  strong atmospheric turbulence channels,'' \emph{Optics express}, vol.~18,
  no.~6, pp. 5356--5366, 2010.

\bibitem{ICC}
M.~Aminikashani, M.~Uysal, and M.~Kavehrad, ``On the performance of mimo fso
  communications over double generalized gamma fading channels,'' \emph{arXiv
  preprint arXiv:1502.00365}, 2015.

\bibitem{24}
A.~Jeffrey and D.~Zwillinger, \emph{Table of integrals, series, and
  products}.\hskip 1em plus 0.5em minus 0.4em\relax Academic Press, 2007.

\bibitem{25}
N.~C. Sagias, G.~K. Karagiannidis, P.~T. Mathiopoulos, and T.~A. Tsiftsis, ``On
  the performance analysis of equal-gain diversity receivers over generalized
  gamma fading channels,'' \emph{IEEE Transactions on Wireless Communications},
  vol.~5, no.~10, pp. 2967--2975, 2006.

\bibitem{29}
V.~Adamchik and O.~Marichev, ``The algorithm for calculating integrals of
  hypergeometric type functions and its realization in reduce system,'' in
  \emph{Proceedings of the international symposium on Symbolic and algebraic
  computation}, 1990, pp. 212--224.

\bibitem{garcia}
A.~Garc{\'\i}a-Zambrana, B.~Castillo-V{\'a}zquez, and C.~Castillo-V{\'a}zquez,
  ``Asymptotic error-rate analysis of fso links using transmit laser selection
  over gamma-gamma atmospheric turbulence channels with pointing errors,''
  \emph{Optics express}, vol.~20, no.~3, pp. 2096--2109, 2012.

\bibitem{wol}
{Wolfram Research Inc.}, ``The wolfram functions site,''
  \url{http://functions.wolfram.com}.

\end{thebibliography}

\end{document}